\documentclass{article}
\usepackage{spconf,amsmath,theorem,graphicx,bm,amssymb,shortcuts_OPT,cite}
\usepackage{microtype}
\usepackage{hyperref,subcaption}
\usepackage{enumitem}
\usepackage{algorithm,algorithmic}
\hypersetup{
  colorlinks   = true, 
  urlcolor     = blue, 
  linkcolor    = blue, 
  citecolor    = red   
}
\usepackage{cleveref}
\usepackage{mdframed}
\newmdtheoremenv{problem}{Problem}
\ninept

\usepackage{pgfplots}
\usetikzlibrary{arrows,shapes,calc,tikzmark,backgrounds,matrix,decorations.markings}
\usepgfplotslibrary{fillbetween}
\pgfplotsset{compat=1.3}

\usepackage{relsize}
\tikzset{fontscale/.style = {font=\relsize{#1}}
    }

\definecolor{lavander}{cmyk}{0,0.48,0,0}
\definecolor{violet}{cmyk}{0.79,0.88,0,0}
\definecolor{burntorange}{cmyk}{0,0.52,1,0}

\definecolor{asuorange}{rgb}{1,0.699,0.0625}
\definecolor{asured}{rgb}{0.598,0,0.199}
\definecolor{asuborder}{rgb}{0.953,0.484,0}
\definecolor{asugrey}{rgb}{0.309,0.332,0.340}
\definecolor{asublue}{rgb}{0,0.555,0.836}
\definecolor{asugold}{rgb}{1,0.777,0.008}

\tikzstyle{stubborn}=[draw,circle, black!80, fill=black!40,
                       text=violet,minimum width=10pt]
\tikzstyle{superpeers}=[draw,circle, asublue!80!white, fill = asublue!50!white,
                       text=violet,minimum width=10pt]
\tikzstyle{susceptible}=[draw,circle, red, fill = red!50,
                       text=violet,minimum width=10pt]
\tikzstyle{latent}=[draw,circle, black!70, fill = black!30,
                       text=violet,minimum width=10pt]
\tikzstyle{legend_general}=[rectangle, rounded corners, thin,
                           black, fill= white, draw, text=black,
                           minimum width=2.5cm, minimum height=0.8cm]
\tikzstyle{legend_graph}=[rectangle, rounded corners, thin,
                           black, fill= white, draw, text=black,
                           minimum width=1cm, minimum height=0.5cm]
\tikzstyle{legend_fw}=[rectangle, rounded corners, very thin,
                           black, fill= asuorange!8, draw, text=black]

\usepackage{color}

\newtheorem{Prop}{Proposition}

\theorembodyfont{\rmfamily}

\title{Spectral partitioning of time-\!varying networks with unobserved edges}
\name{Michael T. Schaub$^{a}$, Santiago Segarra$^{b}$, Hoi-To Wai$^{c}$\thanks{Authors listed in alphabetical order. MTS received funding from the European Union’s Horizon 2020 research and innovation programme under the Marie Sklodowska-Curie grant agreement No 702410. HTW received funding from NSF CCF-BSF 1714672.
The funders had no role in the design of this study; the results presented here reflect solely the authors’ views. 
Emails:~\href{mailto:mschaub@mit.edu}{mschaub@mit.edu}, \href{mailto:segarra@rice.edu}{segarra@rice.edu}, \href{htwai@asu.edu}{htwai@se.cuhk.edu.hk}}}
\address{
    $^{a}$Inst. for Data, Systems and Society, MIT \& Department of Engineering Science, University of Oxford\\ 
    $^{b}$Department of Electrical and Computer Engineering, Rice University\\
    $^{c}$Department of SEEM, The Chinese University of Hong Kong, NT, Hong Kong.
}
\begin{document}
\maketitle
\begin{abstract}
    We discuss a variant of `blind' community detection, in which we aim to partition an unobserved network from the observation of a (dynamical) graph signal defined on the network.
    We consider a scenario where our observed graph signals are obtained by filtering white noise input, and the underlying network is different for every observation.
    In this fashion, the filtered graph signals can be interpreted as defined on a time-varying network. 
    We model each of the underlying network realizations as generated by an independent draw from a latent stochastic blockmodel (SBM).
    To infer the partition of the latent SBM, we propose a simple spectral algorithm for which we provide a theoretical analysis and establish consistency guarantees for the recovery.
    We illustrate our results using numerical experiments on synthetic and real data, highlighting the efficacy of our approach.
\end{abstract}
\begin{keywords}
graph signal processing, topology inference, stochastic blockmodel, community detection, spectral methods
\end{keywords}
\section{Introduction}
\label{sec:intro}
Graph-based tools have become prevalent for the analysis of a range of different systems across the sciences~\cite{Strogatz2001,Newman2010,Jackson2010}.
However, while in many applications we abstract the system under investigation as a network of coupled entities, the underlying couplings are often not known.
Network inference, the problem of determining the interaction topology of a networked system based on a set of nodal observables, has thus gained significant interest over the last years~\cite{Timme2014,Dhaeseleer2000,Brugere2018}.
A number of notions for network inference have featured in the literature, ranging from estimating `functional' couplings based on statistical association measures such as correlation or mutual information~\cite{Friedman2012}, all the way to causal inference~\cite{Pearl2009}.
The notion of inference most pertinent to our discussion is what may be called `topological' inference: given a system of dynamical units, we want to infer their direct physical interactions. 
For example, we would like to infer the adjacency matrix of the network that a distributed system is defined on.
This problem has received wide interest in the literature recently, using techniques from optimization, spectral analysis, and statistics~\cite{Julius2009,Candes2008,Shahrampour2015,Materassi2012,Hayden2016,Yuan2011,wai2016active,Mauroy2017,Segarra2017,giannakis2018topology}.
However, in many situations the goal of inferring the exact network of couplings may be unfeasible for various reasons.
First, we may not have access to a sufficiently large number of samples to fully identify the network.
Second, the network structure itself may be subject to fluctuations over time.
Finally, we may be able to observe only some relevant parts of the system. 

The described challenges need not be fundamental roadblocks since in a number of cases our ultimate target is \emph{not} to obtain the exact network structure.
Rather, our goal is to extract certain mesoscopic features of the network such as important nodes, motifs, or levels of assortativity.
A typical scenario in these lines is the inference of modular structure within the network, i.e., the partitioning of the network into a few blocks, or communities of `similar' nodes according to certain criteria (see~\cite{Fortunato2016,Schaub2017,Abbe2018} for a review on a variety of different approaches).
In this context, the so-called stochastic blockmodel and its related variants~\cite{Abbe2018} have become a major tool for solving this problem from a statistical perspective.
By assuming that the observed network data has been created according to a prescribed generative model, the problem of detecting modular structure is transformed into an estimation problem in which we aim to infer the latent parameters of the model, based on the observed network.

Inspired by our recent work on blind community detection~\cite{Wai2018,Wai2018a,Schaub2018}, in this paper we ask the following question~\cite{Schaub2018}: 

\vspace*{0.05cm}
\noindent \emph{Can we infer the latent partition of a stochastic blockmodel based solely on the observation of a set of nodal signals on the graph without ever observing the underlying graph itself?}
\vspace*{0.2cm}

\noindent\textbf{Contributions and outline} 
We present a fresh look on the network inference problem by advocating an inference approach based on a latent generative model of the network, rather than trying to infer the exact network in terms of its adjacency matrix. 
As we show, this model-based inference procedure that requires only the knowledge of a set of sampled nodal observations can yield surprisingly good results, that are competitive with spectral clustering in which the full network is observed.
We complement the presentation of our blind identification algorithm with a theoretical analysis, in which we we show the statistical consistency of our approach using concentration inequalities and recent results from random matrix theory.

In the remainder of this article, we first discuss our problem setup and associated preliminaries in~\Cref{sec:problem_setup}.
\Cref{sec:main_results} describes our main theoretical results, which underpin our partition inference scheme.
\Cref{sec:numerics} provides numerical illustrations of our results both using synthetic and real-world data.
We conclude with a brief discussion and an outlook on future directions in~\Cref{sec:discussion}.

\section{Problem Formulation}
\label{sec:problem_setup}
\textbf{Graphs, graph signals, and graph filters}.
An~undirected graph $\mathcal G$ consists of a set $\mathcal N$ of $n:=|\mathcal N|$ nodes, and a set $\mathcal E$ of ${n_e:=|\mathcal E|}$ edges, corresponding to unordered pairs of elements in $\mathcal N$.
By identifying the node set $\mathcal N$ with the natural numbers $1,\ldots,n$, such a graph can be compactly encoded by the symmetric adjacency matrix $\bm A$, such that $A_{ij}=A_{ji} = 1$ for all $(i,j)\in\mathcal E$, and $A_{ij} = 0$ otherwise.
Given a graph with adjacency matrix $\bm A$, the (combinatorial) graph Laplacian is defined as $\bm L := \bm D - \bm A$, where $\bm D = \text{diag}(\bm A\bm 1)$ is the diagonal matrix containing the degrees of each node.
We denote the spectral decomposition of the Laplacian by $\bm L = \bm V \bm \Lambda \bm V^\top$.
It is well known that the Laplacian matrix is positive semi-definite~\cite{Merris1994}.

In this paper, we consider filtered signals defined on the graph as described next.
A \emph{graph signal} is a vector $\bm y \in \mathbb{R}^n$ that associates to each node in the graph a scalar-valued observable.
A \emph{graph filter} $\bm{\mathcal{H}}$ of order $T$ is a linear map between graph signals that can be expressed as a matrix polynomial in $\bm L$ of degree $T$
\begin{equation}  
    \bm{\mathcal{H}}(\bm L) = \sum_{k=0}^T h_k \bm L^k.
\end{equation}
Associated with each graph filter, we define the (scalar) generating polynomial $h(\lambda) =  \sum_{k=0}^T h_k \lambda^k$. 
In this work we are concerned with filtered graph signals that can be expressed as
\begin{equation}
    \bm y = \bm{\mathcal{H}}(\bm L) \bm w,
\end{equation}
where $\bm w$ is an excitation signal corresponding to the `initial condition'. We assume that it is zero-mean and white, \ie $\mathbb{E}[\bm w \bm w^\top]= \bm I$, and its entries are bounded almost surely.

Combined with a set of appropriately chosen filter-coefficients, the above signal model can account for a range of interesting signal transformations and dynamics.
This includes consensus dynamics~\cite{Olfati-Saber2007}, random walks and diffusion~\cite{Masuda2017}, as well as more complicated dynamics that can be mediated via interactions commensurate with the graph topology described by the Laplacian \cite{segarra_optimal_2017}.\vspace{.2cm}

\noindent \textbf{Stochastic blockmodel}.
The stochastic blockmodel (SBM) is a latent variable model that defines a probability measure over the set of unweighted networks of fixed size $n$.
In an SBM, the network is assumed to be divided into $k$ groups of nodes.
Each node $i$ in the network is endowed with one latent group label $g_i \in \{1,\ldots,k\}$.
Conditioned on these latent class labels, each link $A_{ij}$ of the adjacency matrix $\bm A \in \{0,1\}^{n\times n}$ is a Bernoulli random variable that takes value $1$ with probability $\Omega_{g_i,g_j}$ and value $0$ otherwise:
\begin{equation}
    A_{ij} | g_i, g_j \sim \text{Ber}(\Omega_{g_i,g_j}).
\end{equation}

To compactly describe the model, we collect all the link probabilities between the different groups in the symmetric affinity matrix $\bm \Omega = [\Omega_{ij}] \in [0,1]^{k \times k}$.
Furthermore we define the partition indicator matrix $\bm G \in \{0,1\}^{n\times k}$ with entries $G_{ij} = 1$ if node $i$ belongs to group $j$ and $G_{ij}=0$ otherwise.
Based on these definitions, we can write the expected adjacency matrix under the SBM as
\begin{equation}\label{E:expected_G}
    \mathbb{E}[\bm A | \bm G] = \bm G \bm \Omega \bm G^\top.
\end{equation}

\noindent \textbf{Observation model and network model inference}.
We observe a nodal signal $\bm y^\tl$ on a network at $m$ instances.
For each instance, we obtain a sample of the form
\begin{equation}\label{eq:signal_model}
    \bm y^{(\ell)} = \bm{\mathcal{H}}(\bm L^{(\ell)}) \bm w^{(\ell)}, \quad \ell = 1,\ldots,m.
\end{equation}
For every $\ell$, we assume that the Laplacian $\bm L^\tl$ is computed from the adjacency matrix of an independently drawn SBM network with a constant parameter matrix $\bm \Omega$.
Moreover, the initial conditions $\bm w^\tl$ are i.i.d. with zero mean and $\EE[ {\bm w}^\tl ( {\bm w}^\tl )^\top ] = {\bm I}$.

Our goal is now to solve the following problem.
\begin{problem}\label{P:main_problem}
    Consider the observation model described by~\Cref{eq:signal_model}.
    Based solely on the $m$ observations $(\bm y^{(1)},\ldots,\bm y^{(m)})$, infer the group structure of the latent SBM generating $\bm L^{(\ell)}$.
\end{problem}

To motivate this setup, consider the example of observing fMRI signals of $m$ different patients in resting state~\cite{damoiseaux2006consistent}.
While for similar patients the overall large-scale structure of each patient's brain network will be similar (the same SBM parameters), the individual details of these networks will be different (each network is a particular realization of the SBM).
Moreover, we do not observe the network itself but only node-measurements ($\bm y^{(\ell)}$), which will generally correspond to different, unknown independent initial conditions ($\bm w^{(\ell)}$).
As a second example, we may think of measuring some node activities such as the expression of opinions at $m$ different, sufficiently separated instances of time in some form of social network. 
Assuming a reasonable stable social fabric, the large scale features of the latent (unobserved) network should be relatively stable, while the individual active links in each observation instance may be different.

\section{Algorithm and Theoretical Analysis}
\label{sec:main_results}
\Cref{alg:timevary} describes a simple spectral method to solve Problem~\ref{P:main_problem}. 
In a nutshell, given the observations $\{\bm y^\tl\}_{\ell=1}^m$, we compute their sample covariance $\widehat{\bm C}_y^m$ as in \eqref{eq:cov} and then apply $k$-means clustering on the leading eigenvectors of $\widehat{\bm C}_y^m$.
For simplicity, we assume here that the number of groups $k$ of the SBM is known.
However, $k$ could be estimated as well from the spectral properties of the covariance matrix, e.g., by estimating its effective rank.

To theoretically assess the performance of the proposed method, we present an analysis in three steps. 
First, we characterize the rate of convergence of the sample covariance to the true covariance ${\bm C}_y$ (cf. Proposition~\ref{prop:conc}). 
Second, we determine the structure of the limiting matrix ${\bm C}_y$ (cf. Proposition~\ref{P:limit_convergence}). 
Finally, we show that the eigenstructure of ${\bm C}_y$ contains all the information needed to solve Problem~\ref{P:main_problem} (cf. Proposition~\ref{P:recovery}).

\algsetup{indent=1em}
\begin{algorithm}[tb]
	\caption{Spectral partitioning of time-varying networks.}\label{alg:timevary}
	\begin{algorithmic}[1]
	\STATE \textbf{Input}: filtered graph signals $\{\bm y^\tl\}_{\ell=1}^m$; number of groups $k$.
        \STATE Compute the sampled covariance matrix as\vspace{-.1cm} 
		\beq \label{eq:cov}
		\textstyle \widehat{\bm C}_y^m \eqdef (1/m) \sum_{\ell=1}^m ({\bm y}^\tl ) ( {\bm y}^\tl )^\top \vspace{-.4cm}
		\eeq
		\STATE Evaluate the EVD of $\widehat{\bm C}_y^m$
		as $\widehat{\bm C}_y^m = \widehat{\bm V} \widehat{\bm{\Lambda}} \widehat{\bm V}^\top$.
		\STATE Apply $k$-means on the row vectors of the matrix 
		$\widehat{\bm V}_k \in \RR^{n \times k}$, whose columns are
		the top-$k$ eigenvectors in $\widehat{\bm C}_y^m$. 
	\STATE \textbf{Output}: a partition ${\cal N} = {\cal C}_1 \cup ... \cup {\cal C}_k$ with ${\cal C}_i \cap {\cal C}_j = \emptyset$ if $i \neq j$.
	\end{algorithmic} 
\end{algorithm}

Recall the definition of the covariance matrix
\beq\label{eq:exp_cov}
{\bm C}_y \eqdef \EE[ {\bm y}^\tl ( {\bm y}^\tl )^\top ],
\eeq
where the expected value is taken over both sources of randomness, i.e., the excitation signal ${\bm w}^\tl$ as well as the Laplacian $\bm L^{(\ell)}$ of the realized graph. Based on this, the following result can be shown.

\begin{Prop}  \label{prop:conc}
	Assume that the following conditions hold:
	\begin{enumerate}[itemsep=0.1mm]
	\item[(a)] The spectral norm of the graph filter is uniformly bounded, i.e., $\|\bm{\mathcal H} (\bm L^{(\ell)})\|_2 \leq \bar{h}$ for all $\ell$.
	\item[(b)] The excitation signal satisfies $\| {\bm w}^\tl \|_2 \leq c \sqrt{n}$ 
	almost surely, and $( \EE [ \| {\bm w}^\tl \|_2^q ] )^{1/q} \leq W_0 < \infty$ for some $q \geq 4$. 
	\end{enumerate}
	Then, for any $m \geq n \geq 4$, with probability at least $1-\delta$, 
	one has
	\beq \label{eq:result1}
	\big\| \widehat{\bm C}_y^m - {\bm C}_y \big\|_2 \leq c_0 \!~ ( \log \log n )^2 \left( \frac{n}{m} \right)^{\frac{1}{2}-\frac{2}{q}}\eqs,
	\eeq
	where the constant $c_0$ depends on $q$, $\bar{h}$, $\delta$, and $W_0$. 
\end{Prop}
\noindent\textbf{Proof.}  
Observe that the following bound
\beq \notag
\| {\bm y}^{(\ell)} \|_2 = \| {\cal H} ( {\bm L}^{(\ell)} ) {\bm w}^{(\ell)} \|_2
\leq  \| {\cal H} ( {\bm L}^{(\ell)} ) \|_2 \!~ \| {\bm w}^{(\ell)} \|_2,
\eeq
combined with condition \emph{(a)} implies that
\beq \label{eq:l2bound}
\| {\bm y}^{(\ell)} \|_2 \leq \bar{h} \!~  \| {\bm w}^{(\ell)} \|_2 \eqs.
\eeq
To show that $\widehat{\bm C}_y^m$ converges to its expected value, 
first we observe 
from \eqref{eq:l2bound} that
\beq \label{eq:cond1}
\| {\bm y}^{(\ell)} \|_2 \leq c \bar{h} \!~   \sqrt{n}~~a.s. \eqs,
\eeq
if $\| {\bm w}^{(\ell)} \|_2 \leq c \sqrt{n}$ for some $c$ almost surely. 
Second, 
consider any ${\bm u}$ such that $\| {\bm u} \|_2 = 1$, 
we have 
\beq \label{eq:keyineq}
| \langle {\bm y}^{(\ell)}, {\bm u} \rangle | \leq \| {\bm y}^{(\ell)} \|_2 \| {\bm u} \|_2 
\leq \bar{h}\!~ \| {\bm w}^{(\ell)} \|_2 \eqs.
\eeq
Applying \eqref{eq:keyineq}, for any $q \geq 1$, one has
\beq \label{eq:cond2}
\begin{split}
	\big( \EE [ | \langle {\bm y}^{(\ell)}, {\bm u} \rangle |^q ] \big)^{1/q}
	& \leq \bar{h}  \!~ (\EE [ \| {\bm w}^{(\ell)} \|_2^q ])^{1/q} \leq \bar{h} \!~ W_0 \eqs.
\end{split}
\eeq
From \eqref{eq:cond1} and \eqref{eq:cond2}, the two conditions in \cite[Eq.~(2.2)]{vershynin12} hold. 
Invoking \cite[Theorem 6.1]{vershynin12} shows 
the desired result in \eqref{eq:result1}. \hfill
$\blacksquare$
\vspace{1mm}

The conditions required by the proposition are mild. For instance, condition (a) holds
for graph filters that are \emph{low-pass} \cite{Wai2018}. Indeed, in such a case we have that $\|\bm{\mathcal H} (\bm L^{(\ell)})\|_2 \leq h(0)$, where $h(\cdot)$ is the generating polynomial of the filter $\bm{\mathcal H}(\cdot)$.
Condition (b) holds with for $q \geq 4$ 
when the excitation signal is bounded, e.g.,
$w_i^\tl$ is i.i.d.~and distributed with ${\cal U}[-b,b]$, $b < \infty$. 
The proposition shows that the sampled covariance converges
to the true covariance at a rate ${\cal O}( 1 / m^{\frac{1}{2} - \frac{2}{q}} )$.
In particular, the convergence rate is ${\cal O}( \sqrt{1/m} )$
in the case of bounded excitation signals, where $q$ can be made arbitrarily large.

Notice that Proposition~\ref{prop:conc} concerns general covariance matrices and does not use the fact that $\bm L^{(\ell)}$ is the Laplacian of a graph drawn from an SBM. 
In order to derive results about the recovery of the latent communities, we will have to put this assumption into place.
For simplicity, we consider in the theoretical considerations that follow a simple planted partition model of size $n$, in which only two equally sized communities of size $n/2$ exist~\cite{Abbe2018}.
Nonetheless, the arguments that follow can be extended to general SBMs.

In our planted partition model, the probability of an edge between two nodes within the same community is governed by the parameter $a$ whereas the probability of a link between two nodes of different communities is described by parameter $b$. 
Given two nodes $i$ and $j$, the expression $i \sim j$ denotes that both nodes lie in the same block of the SBM, whereas $i \not\sim j$ indicates the contrary. 
Moreover, for simplicity we denote by $\bm H =\bm{\mathcal{H}}( {\bm L}^\tl )$ the (random) matrix representing the filter of interest. We use the following parameters to denote the expected entries of $\bm H$:
\begin{align}\label{E:parameters_p_1_8}
&p_1 := \EE[H_{ii}^2] \,\, \text{for all} \,\, i , \,\, &&p_2 := \EE[H_{ij}^2] \,\,\text{for} \,\, i \sim j, \nonumber\\
&p_3 := \EE[H_{ij}^2] \,\,\text{for} \,\, i \not\sim j, \,\, &&p_4 := \EE[H_{il} H_{jl}] \,\,\text{for} \,\, i \sim j \sim l, \nonumber\\
&p_5 := \EE[H_{ii} H_{ji}] \,\,\text{for} \,\, i \sim j, \,\, &&p_6 := \EE[H_{il} H_{jl}] \,\,\text{for} \,\, i \sim j \not\sim l, \nonumber\\
&p_7 := \EE[H_{ii} H_{ji}] \,\,\text{for} \,\, i \not\sim j, \,\, &&p_8 := \EE[H_{il} H_{jl}] \,\,\text{for} \,\, i \not\sim j. \nonumber
\end{align}

Based on the introduced notation, we characterize the covariance structure of our observed output signals.

\begin{Prop}\label{P:limit_convergence}
	The covariance ${\bm C}_y$ defined in \eqref{eq:exp_cov} is given by
	\begin{equation}\label{E:expected_value}
	{\bm C}_y = (c_3 - c_1) \bm I + \bm G 
	\begin{pmatrix}
	c_1 & c_2 \\
	c_2 & c_1
	\end{pmatrix}
	\bm G ^\top,
	\end{equation}
	where $\bm G \in \{0,1\}^{n\times2}$ is the partition indicator matrix as defined before \eqref{E:expected_G}, and the constants $c_i$ are given by $c_1 = (\frac{n}{2}-2) p_4 + 2 p_5 + \frac{n}{2} p_6$, $c_2 = 2 (\frac{n}{2}-1) p_8 + 2 p_7$, and $c_3 = p_1 + (\frac{n}{2}-1)p_2 + \frac{n}{2}p_3$.
\end{Prop}

\noindent\textbf{Proof.} 
Consider first the diagonal entries of ${\bm C}_y$, we have that
\begin{align}
\textstyle
[{\bm C}_y]_{ii} &= \EE[\bm h_i^\top \bm w \bm w^\top \bm h_i]= \EE\Big[ \Big(\sum_j H_{ij} w_j\Big)^2 \Big] \nonumber \\
&= \EE\Big[\sum_j H_{ij}^2 w_j^2 + \sum_{j,k} H_{ij} w_j H_{ik} w_i\Big] \nonumber \\
&= \sum_j \EE[H_{ij}^2] \EE[w_j^2] + \sum_{j,k} \EE[H_{ij} H_{ik}] \EE[w_j] \EE[w_i] \nonumber
\end{align}
Using the fact that $\EE[w_j^2] = 1$ and $\EE[w_j]=0$, it follows that
\begin{align}\label{E:proof_diag_element}
[{\bm C}_y]_{ii} &= \EE[H_{ii}^2] + \sum_{j \sim i} \EE[H_{ij}^2] + \sum_{j \not\sim i} \EE[H_{ij}^2] \\
&= p_1 + \Big(\frac{n}{2}-1\Big) p_2 + \frac{n}{2} p_3 = c_3. \nonumber
\end{align}
Next, we consider an off-diagonal entry in ${\bm C}_y$ within a block of the SBM, i.e., for $i \sim j$ but $i \neq j$ we have that
\begin{align}
\textstyle
[{\bm C}_y]_{ij} &= \EE[\bm h_i^\top \bm w \bm w^\top \bm h_j]= \EE\Big[ \sum_{l,k} H_{il} w_l H_{jk} w_k\Big] \nonumber \\
& \overset{(a)}{=} \EE\Big[ \sum_l H_{il} H_{jl} w_l^2 \Big] \overset{(b)}{=} \sum_l \EE[H_{il} H_{jl}] \nonumber,
\end{align}
where (a) follows from $\EE[w_l w_k] = 0$ whenever $l \neq k$, and (b) used that $\EE[w_l^2]=1$. From the above it then follows that
\begin{align}\label{E:proof_off_diag_element_1}
[{\bm C}_y]_{ij} &= 2 \EE[H_{ii}H_{ji}] + \!\!\! \sum_{l | l \sim i, j \neq l \neq i} \!\!\! \EE[H_{il} H_{jl}] + \sum_{l | l \not\sim i} \EE[H_{il} H_{jl}] \nonumber \\
&= 2 p_5 + \Big( \frac{n}{2} - 2 \Big) p_4 + \frac{n}{2} p_6 = c_1.
\end{align}
Finally, considering $i$ and $j$ in different blocks, we can similarly show that $[{\bm C}_y]_{ij} = c_2$. By combining this result with \eqref{E:proof_diag_element} and \eqref{E:proof_off_diag_element_1}, expression \eqref{E:expected_value} readily follows.
\hfill
$\blacksquare$

\vspace{1mm}

An important consequence of~\Cref{P:limit_convergence} is the resulting spectral decomposition of ${\bm C}_y$ and how this eigenstructure relates to the planted (true) communities in the underlying SBM. 
The following proposition combines the results from Propositions~\ref{prop:conc} and \ref{P:limit_convergence} and justifies (asymptotically) the performance of Algorithm~\ref{alg:timevary} in recovering the true communities.

\begin{Prop}\label{P:recovery}
    Assume that the conditions in Proposition~\ref{prop:conc} hold, and that $c_1 > |c_2|$, as defined in Proposition~\ref{P:limit_convergence}. Then, for a large enough number of observations $m$, Algorithm~\ref{alg:timevary} is guaranteed to recover the two communities of the equisized planted partition model.
\end{Prop}	
\noindent\textbf{Proof.} 
Direct computation from expression~\eqref{E:expected_value} reveals that the vector of all ones $\mathbf{1}$ is an eigenvector of ${\bm C}_y$ with associated eigenvalue $ \mu_1 := \frac{n}{2} (c_1+c_2) + (c_3-c_1)$. 
Similarly, the signed binary vector $\pm \mathbf{1} := {\bm G} [1,-1]^\top$ whose sign indicates membership to each community is also an eigenvector of ${\bm C}_y$ but with eigenvalue $\mu_2 := \frac{n}{2} (c_1-c_2) + (c_3-c_1)$. 
Every other eigenvector is associated with the eigenvalue $\mu := c_3 - c_1$. 
Given that Algorithm~\ref{alg:timevary} keeps the top-$2$ eigenvectors of $\widehat{\bm C}_y^m$, it follows from the concentration result in Proposition~\ref{prop:conc} that whenever $\mu_1 > \mu$ and $\mu_2 > \mu$, the eigenvectors selected by our algorithm will be arbitrarily close to $\mathbf{1}$ and $\pm \mathbf{1}$ for large enough $m$, thus leading to perfect recovery. 
Hence, we need $c_1+c_2 >0$ and $c_1-c_2 >0$, from where $c_1 > |c_2|$ follows.
\hfill
$\blacksquare$
\vspace{1mm}

The constants $c_1$ and $c_2$ depend on the parameters $p_4$ through $p_8$, which in turn depend on the filter specification $h(\cdot)$ and the probabilities $a$ and $b$ in the considered SBM. 
Whenever $a=b$, it can be shown that $c_1=c_2$, thus preventing the recovery of the planted true communities, as expected.
Given a generic filter for which $c_1 > |c_2|$ if $a\neq b$, however, even a minimal difference between $a$ and $b$ will result asymptotically in a perfect recovery.
This is in contrast with the detectability limit that holds for the SBM recovery problem with an observed network, where the partitions cannot be recovered if $a$ is too close to $b$~\cite{Abbe2018}. 
The reason behind the improved resolution here is that in our problem each sample $\bm y^\tl$ corresponds to an (indirect) observations of a \emph{different} graph drawn from the same SBM, allowing us to detect communities for large enough samples $m$ even in the most adverse scenarios.
When inferring an SBM from a single network observation, one cannot (indirectly) leverage such additional graph samples, resulting in a detectability limit~\cite{Abbe2018}.

\section{Numerical Experiments}\label{sec:numerics}

\noindent \textbf{Synthetic data}. 
We first examine the claims made in the paper using synthetic data. 
We draw graphs from an SBM with $n=100$ nodes and $k=2$ communities, 
with $\Omega_{g_i,g_j} = 4 \log n / n$ if $g_i = g_j$, and
$\Omega_{g_i,g_j} = 4 \gamma \log n / n$ otherwise, parametrized by $\gamma \in (0,1)$. 
Note that the smaller $\gamma$ is, the easier it is to detect the communities. 
Throughout the section, the input signal is i.i.d.~and set as ${\bm w}^\tl \sim {\cal U}[-1,1]^n$.
The graph filter considered is 
${\cal H}( {\bm L} ) = ({\bm I} - \alpha {\bm L}^\tl )^5$ where
$\alpha = 1 / (4 + 4\gamma) \log n$ ensures that 
$\|\bm{\mathcal H}(\bm{L}^{(\ell)})\|<1$ for all $\ell$. 

\begin{figure}[tb!]
\centering
\resizebox{.95\linewidth}{!}{\resizebox{.9\linewidth}{!}
{ \sf 
\begin{tikzpicture}
\pgfplotsset{ scale only axis,
    width=0.55\textwidth,height=0.3\textwidth,
    grid=both,grid style={line width=.01pt, draw=gray!30},
    legend cell align=left, legend style={legend pos=south west,font=\large},
    xlabel={\large Sample size $m$},
    enlarge x limits=0.05,
    enlarge y limits=0.05, ymin = 0.5e-4,   xmin = 10
}

\pgfplotsset{every tick label/.append style={font=\large}}

\begin{loglogaxis}[ylabel={\large Error Rate}]
\addplot[asublue, very thick, mark options={solid,mark size=3}, mark=*] 
      table[x index=0, y index=1, col sep=comma] {./ResLim_01a_err_CYnl.csv};
      \addlegendentry{$\gamma = 0.1$};
\addplot[asured, very thick, mark options={solid,mark size=3}, mark=square] 
      table[x index=0, y index=1, col sep=comma] {./ResLim_03a_err_CYnl.csv};
      \addlegendentry{$\gamma = 0.3$};
\addplot[asured!50!blue, very thick, mark options={solid,mark size=3}, mark=triangle] 
      table[x index=0, y index=1, col sep=comma] {./ResLim_05a_err_CYnl.csv};
      \addlegendentry{$\gamma = 0.5$};
\addplot[green!50!black, very thick, mark options={solid,mark size=3}, mark=o] 
      table[x index=0, y index=1, col sep=comma] {./ResLim_07a_err_CYnl.csv};
      \addlegendentry{$\gamma = 0.7$};
\addplot[asuorange, very thick, mark options={solid,mark size=3}, mark=diamond] 
      table[x index=0, y index=1, col sep=comma] {./ResLim_09a_err_CYnl.csv};
      \addlegendentry{$\gamma = 0.9$};
\addplot[black, dashed, very thick] 
      table[x index=0, y index=1, col sep=comma] {./ResLim_01a_err_S.csv};
      \addlegendentry{SC ($\gamma = 0.1$)};
\end{loglogaxis}

\end{tikzpicture}
} }\vspace{-.2cm}
\caption{Error rate of partition recovery using~\Cref{alg:timevary} against sample size for synthetic 
time varying graphs with various structural parameters $\gamma$ (see text).} \label{fig:reslim}
\end{figure}
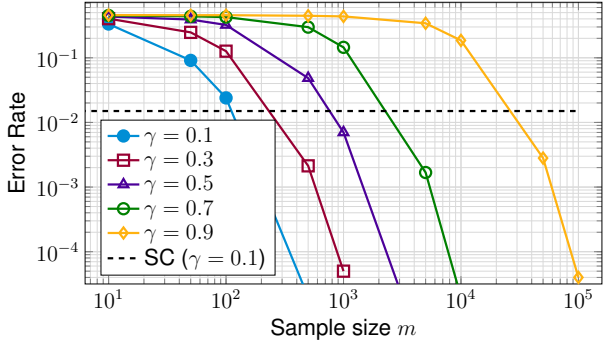

In Fig.~\ref{fig:reslim} we simulate the error rate of the partition inference over different settings of $\gamma$, 
against the sample size $m$ using our proposed method. We found that the 
error rate decays to zero asymptotically as $m \rightarrow \infty$ regardless of the 
connectivity probability parameter $\gamma$. Moreover, the error rate
is markedly better compared to the application of standard spectral clustering ({\sf SC}) on a single
instance of the graph Laplacian.
Note that this holds even if the graph considered for {\sf SC} is taken from an SBM with $\gamma = 0.1$, in line with our discussion at the end of~\Cref{sec:main_results}.

\noindent \textbf{United States Senate data}. We apply the proposed 
method to rollcall data (available at \texttt{https://voteview.com})
taken from the 110th to 114th congress of the US Senate
(corresponding to years 2007 to 2017) consisting of $m=2998$ rollcalls.
Using this data we focus on inferring partitions of a network in which the nodes represent the $n=50$ states of USA.
To convert the data into real-valued graph signals that agree with our time varying topology model, the $\ell$th rollcall data is mapped into a sample graph signal ${\bm y}^\ell \in \RR^{50}$ as follows. 
For each state $i \in \{1,...,50\}$, we compute $y_i^\ell \in [-1,1]$ as the average vote value from the 
two senators of each state, where the vote value counts a `Yay' as $1$, an absentee or an abstain as $0$, and a `Nay' as $-1$. 
Note that with the framework of our model, we assume that the \emph{community} a state belongs to remains \emph{invariant} since the economic/political situation of the state varies slowly in general, even though senators maybe elected in/out during different periods.

\begin{figure}[t]
\centering
    \begin{subfigure}{0.475\linewidth}
\includegraphics[width=0.99\linewidth]{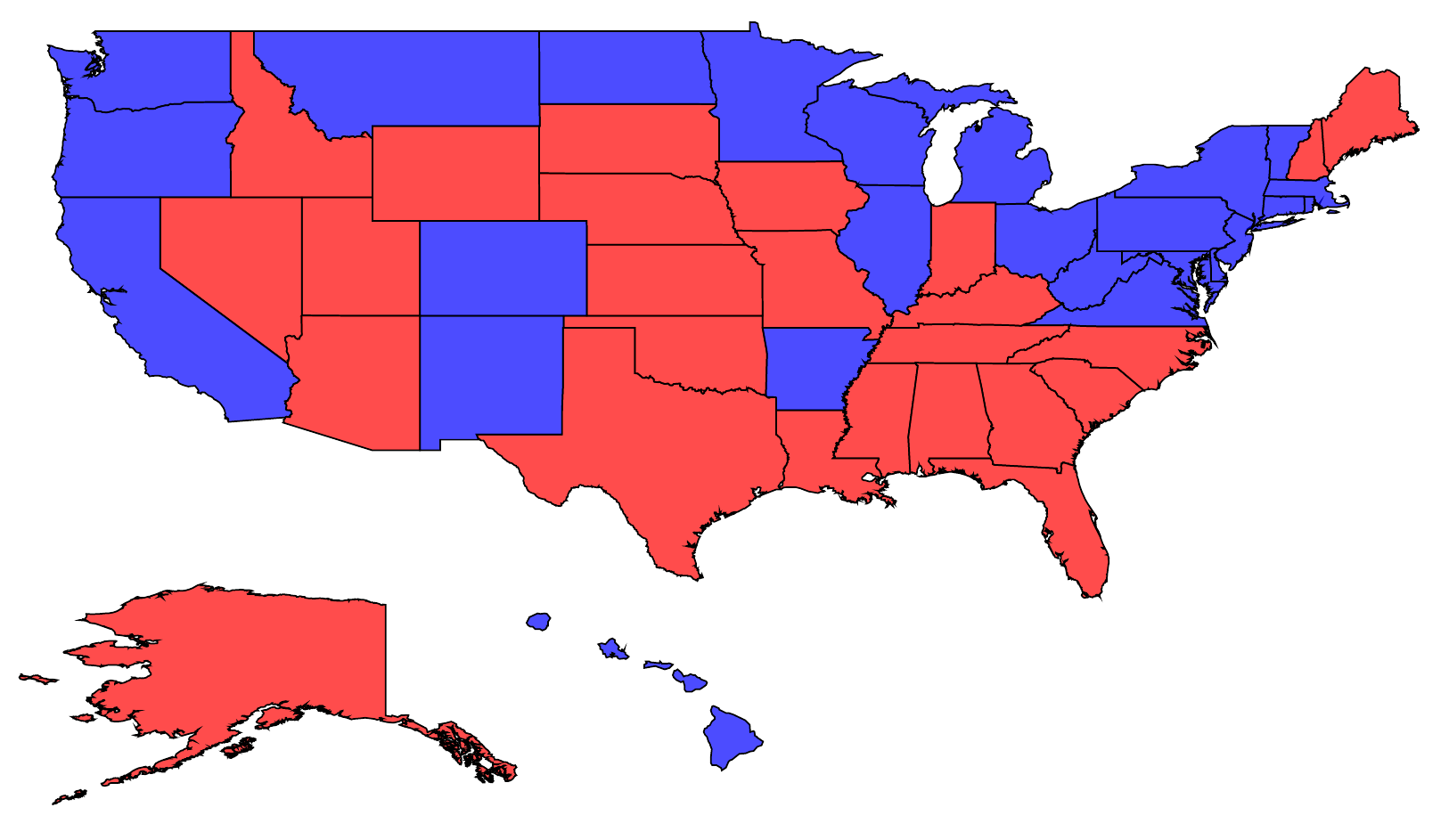}
\caption{$k\!=\!2$. Data from 2007-2017.}
    \end{subfigure}
    \begin{subfigure}{0.475\linewidth}
    \includegraphics[width=.99\linewidth]{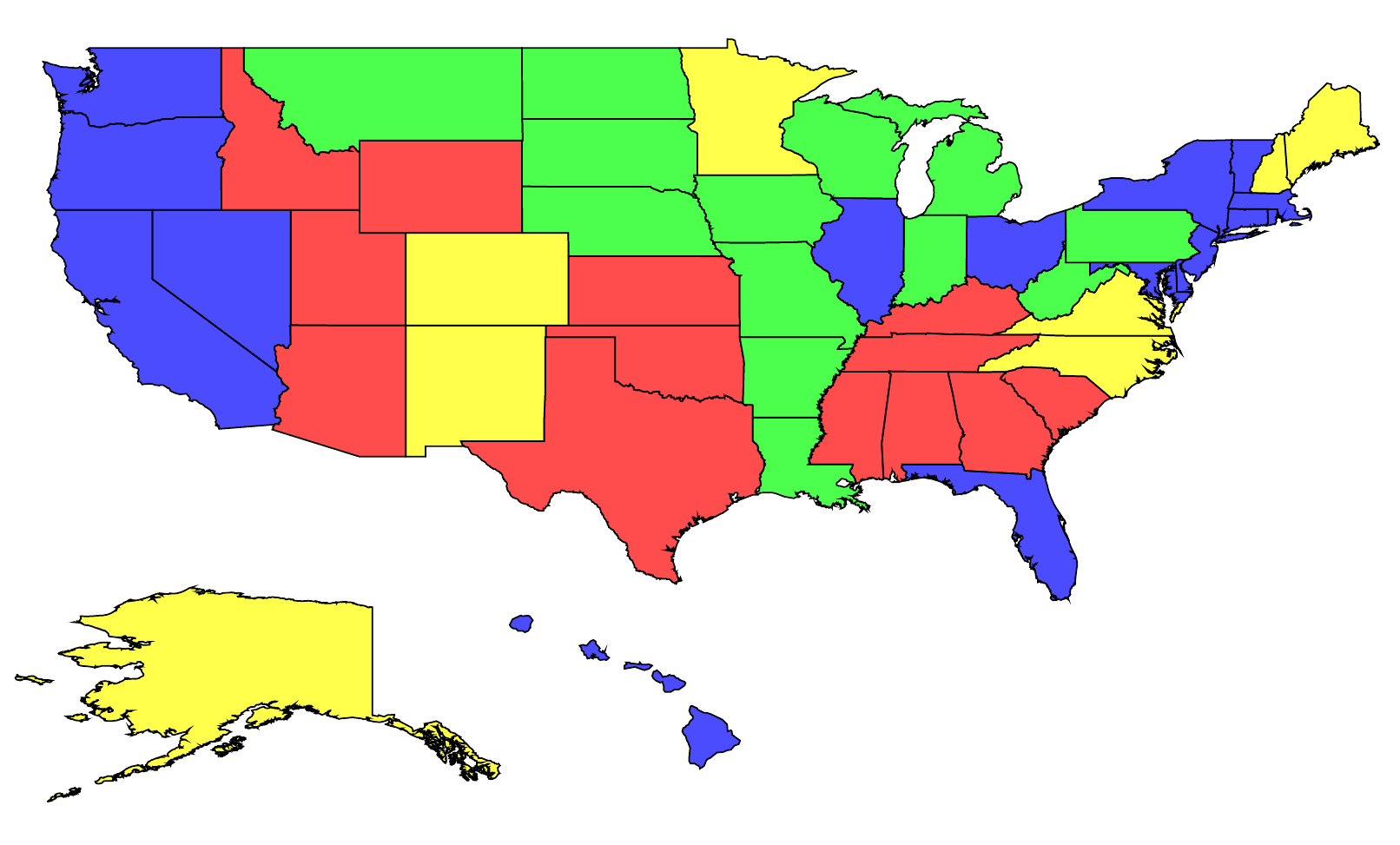}\vspace{-.15cm}
    \caption{$k\!=\!4$. Data from 2007-2017.}
    \end{subfigure}
        \begin{subfigure}{0.475\linewidth}
    \includegraphics[width=.99\linewidth]{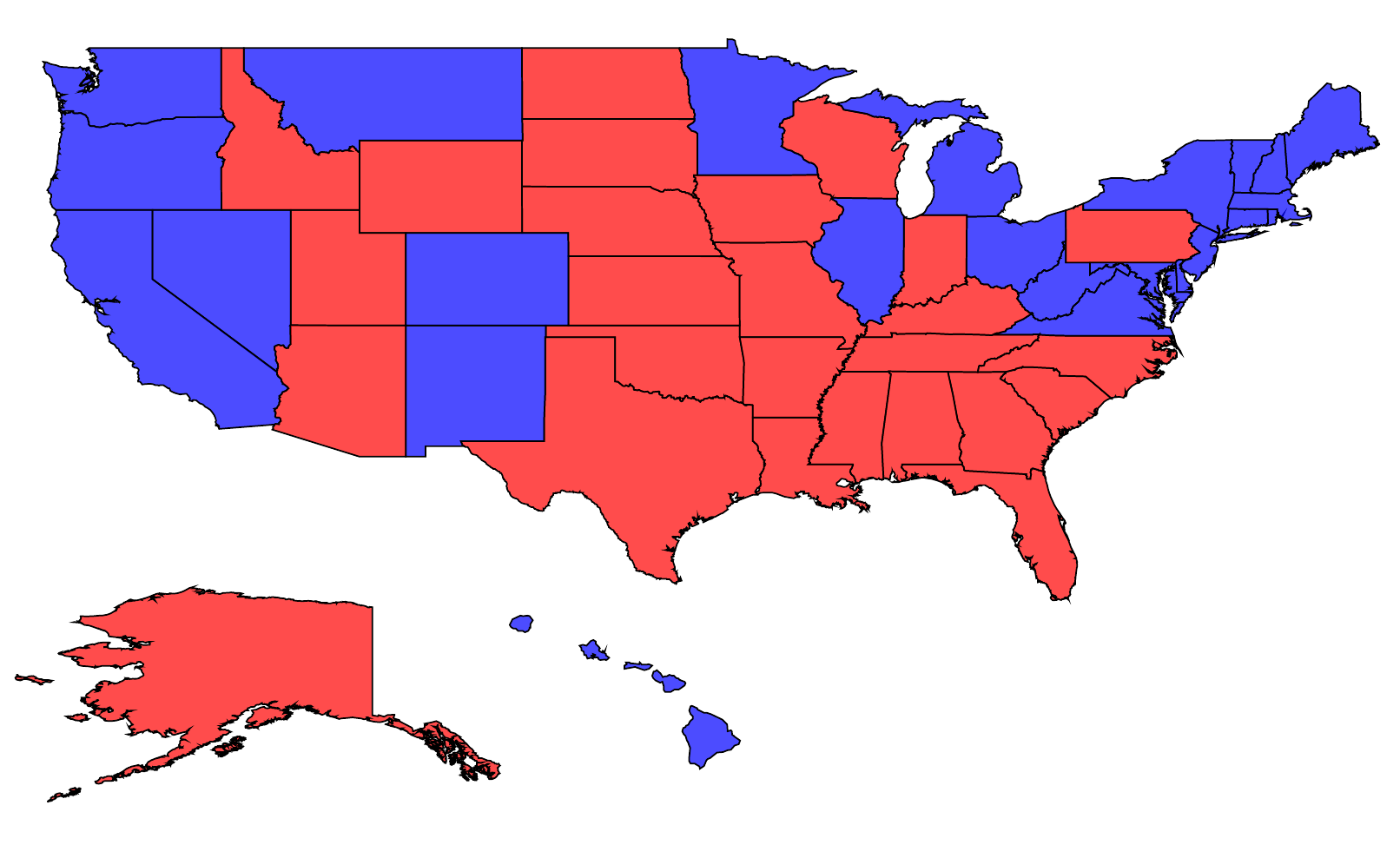}
    \caption{$k\!=\!2$. Data from 2015-2017.}
    \end{subfigure}
    \begin{subfigure}{0.475\linewidth}
\includegraphics[width=.99\linewidth]{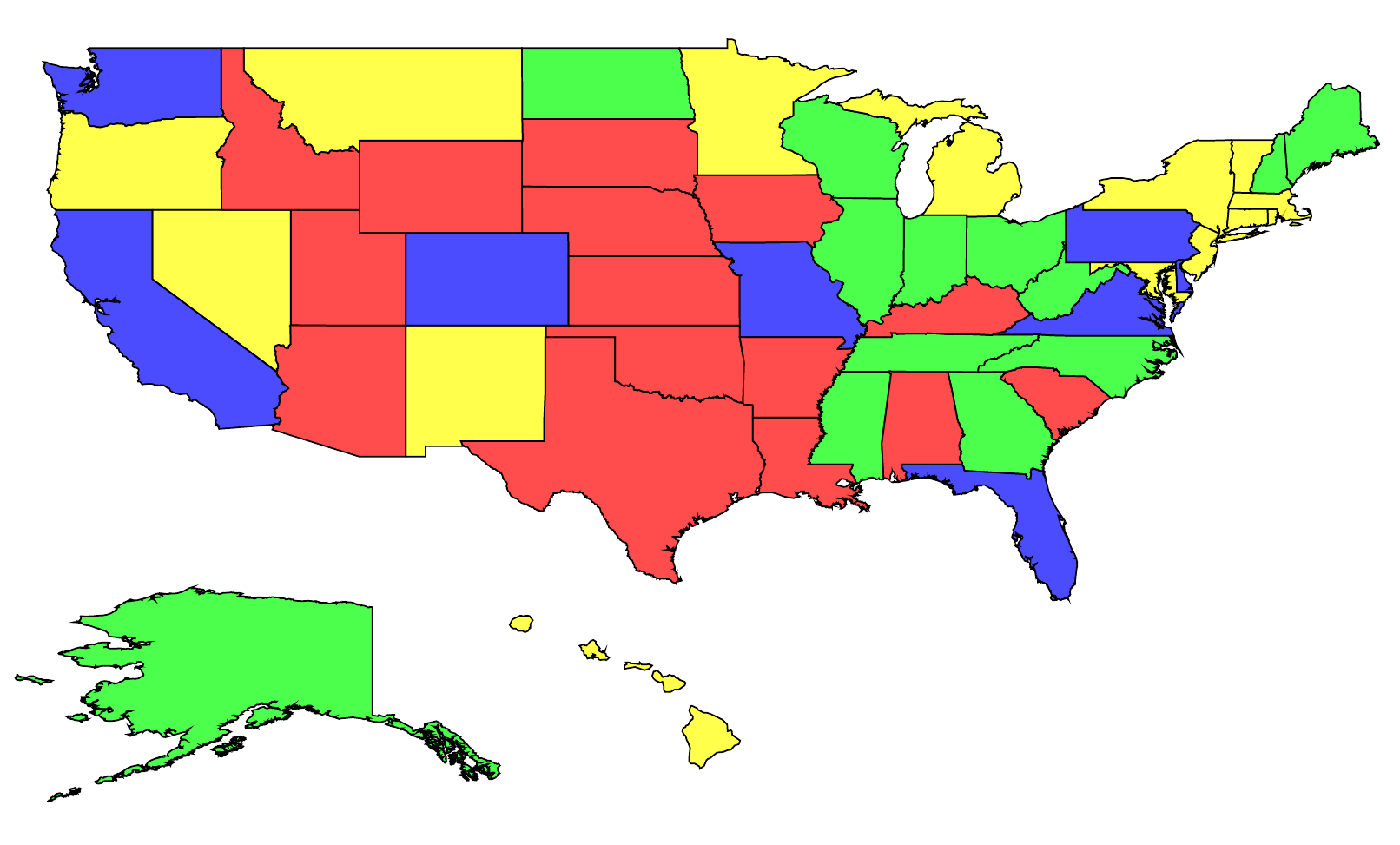}
\caption{$k\!=\!4$. Data from 2015-2017.}
\end{subfigure}\vspace{-.1cm}
\caption{Partitioning of the US Senate's states network for different number of communities $k$ and observation periods (see text).} \label{fig:senate}\vspace{-.3cm}
\end{figure}

Fig.~\ref{fig:senate} shows the partitions of the states 
at different resolution ($k=2,4$) based on the rollcall data from the combined 
periods of 2007-2017 (Fig.~\ref{fig:senate}a,b) and from the latest period 2015-2017 (Fig.~\ref{fig:senate}c,d), respectively. 
At a resolution of $k=2$, the partition result corroborates the common belief 
about the division between `Republican' (red, e.g., Texas \& Arizona) and `Democrat' (blue, e.g., California \& Massachusetts) states, with the 2015-2017 data reflecting recent changes in the elected senators for states such as Maine and New Hampshire.
We also remark that for $k=4$, the partitioning result using 
2015-2017 data is less conclusive as it changes substantially when we sample 
a small batch of rollcall data. Such instability is not observed in the 2007-2017 data at the same resolution, where the partition identifies some of the `swing' states such as Michigan and Louisiana.

\section{Discussion}\label{sec:discussion}
Network inference is often a critical step to perform any kind of network analysis.
In certain cases, however, we are only interested in extracting some coarser features of the network, e.g., in the form of communities~\cite{Wai2018,Wai2018a,Schaub2018,Hoffmann2018}.
As we have shown in this manuscript, if we have access to a set of independent samples from a filtered signal defined on the nodes of the network, this task can be achieved even in the absence of any information about the edges.
As we have discussed for the system studied here, if the underlying network is time-varying but its latent structure remains stationary, we may even obtain a better partition recovery performance when compared to observing a single full snapshot of the actual network.
Characterizing this trade-off and the sample complexity of the corresponding problems in more detail, as well as enlarging the class of latent models and considered graph filters are interesting avenues for future work.

\newpage
\label{sec:refs}
\bibliographystyle{IEEEbib}
\bibliography{references}

\begin{thebibliography}{10}

\bibitem{Strogatz2001}
Steven~H. Strogatz,
\newblock ``{E}xploring complex networks,''
\newblock {\em Nature}, vol. 410, no. 6825, pp. 268--276, Mar. 2001.

\bibitem{Newman2010}
Mark E.~J. Newman,
\newblock {\em {N}etworks: {A}n {I}ntroduction},
\newblock Oxford University Press, USA, Mar. 2010.

\bibitem{Jackson2010}
Matthew~O Jackson,
\newblock {\em Social and economic networks},
\newblock Princeton university press, 2010.

\bibitem{Timme2014}
Marc Timme and Jose Casadiego,
\newblock ``Revealing networks from dynamics: an introduction,''
\newblock {\em Journal of Physics A: Mathematical and Theoretical}, vol. 47,
  no. 34, pp. 343001, 2014.

\bibitem{Dhaeseleer2000}
Patrik D'haeseleer, Shoudan Liang, and Roland Somogyi,
\newblock ``Genetic network inference: from co-expression clustering to reverse
  engineering,''
\newblock {\em Bioinformatics}, vol. 16, no. 8, pp. 707--726, 2000.

\bibitem{Brugere2018}
Ivan Brugere, Brian Gallagher, and Tanya~Y Berger-Wolf,
\newblock ``Network structure inference, a survey: Motivations, methods, and
  applications,''
\newblock {\em ACM Computing Surveys (CSUR)}, vol. 51, no. 2, pp. 24, 2018.

\bibitem{Friedman2012}
Jonathan Friedman and Eric~J Alm,
\newblock ``Inferring correlation networks from genomic survey data,''
\newblock {\em PLoS computational biology}, vol. 8, no. 9, pp. e1002687, 2012.

\bibitem{Pearl2009}
Judea Pearl,
\newblock ``Causal inference in statistics: An overview,''
\newblock {\em Statistics surveys}, vol. 3, pp. 96--146, 2009.

\bibitem{Julius2009}
Agung Julius, Michael Zavlanos, Stephen Boyd, and George~J Pappas,
\newblock ``Genetic network identification using convex programming,''
\newblock {\em IET systems biology}, vol. 3, no. 3, pp. 155--166, 2009.

\bibitem{Candes2008}
Emmanuel~J Candes, Michael~B Wakin, and Stephen~P Boyd,
\newblock ``Enhancing sparsity by reweighted l1 minimization,''
\newblock {\em Journal of Fourier analysis and applications}, vol. 14, no. 5-6,
  pp. 877--905, 2008.

\bibitem{Shahrampour2015}
Shahin Shahrampour and Victor~M Preciado,
\newblock ``Topology identification of directed dynamical networks via power
  spectral analysis,''
\newblock {\em IEEE Transactions on Automatic Control}, vol. 60, no. 8, pp.
  2261, 2015.

\bibitem{Materassi2012}
Donatello Materassi and Murti~V Salapaka,
\newblock ``On the problem of reconstructing an unknown topology via locality
  properties of the wiener filter,''
\newblock {\em IEEE transactions on automatic control}, vol. 57, no. 7, pp.
  1765--1777, 2012.

\bibitem{Hayden2016}
David Hayden, Young~Hwan Chang, Jorge Goncalves, and Claire~J Tomlin,
\newblock ``Sparse network identifiability via compressed sensing,''
\newblock {\em Automatica}, vol. 68, pp. 9--17, 2016.

\bibitem{Yuan2011}
Ye~Yuan, Guy-Bart Stan, Sean Warnick, and Jorge Goncalves,
\newblock ``Robust dynamical network structure reconstruction,''
\newblock {\em Automatica}, vol. 47, no. 6, pp. 1230--1235, 2011.

\bibitem{wai2016active}
Hoi-To Wai, Anna Scaglione, and Amir Leshem,
\newblock ``Active sensing of social networks,''
\newblock {\em IEEE Transactions on Signal and Information Processing over
  Networks}, vol. 2, no. 3, pp. 406--419, 2016.

\bibitem{Mauroy2017}
Alexandre Mauroy and Julien Hendrickx,
\newblock ``Spectral identification of networks using sparse measurements,''
\newblock {\em SIAM Journal on Applied Dynamical Systems}, vol. 16, no. 1, pp.
  479--513, 2017.

\bibitem{Segarra2017}
Santiago Segarra, Antonio~G Marques, Gonzalo Mateos, and Alejandro Ribeiro,
\newblock ``Network topology inference from spectral templates,''
\newblock {\em IEEE Transactions on Signal and Information Processing over
  Networks}, vol. 3, no. 3, pp. 467--483, 2017.

\bibitem{giannakis2018topology}
Georgios~B Giannakis, Yanning Shen, and Georgios~Vasileios Karanikolas,
\newblock ``Topology identification and learning over graphs: Accounting for
  nonlinearities and dynamics,''
\newblock {\em Proceedings of the IEEE}, vol. 106, no. 5, pp. 787--807, 2018.

\bibitem{Fortunato2016}
Santo Fortunato and Darko Hric,
\newblock ``Community detection in networks: A user guide,''
\newblock {\em Physics Reports}, vol. 659, pp. 1--44, 2016.

\bibitem{Schaub2017}
Michael~T Schaub, Jean-Charles Delvenne, Martin Rosvall, and Renaud Lambiotte,
\newblock ``The many facets of community detection in complex networks,''
\newblock {\em Applied Network Science}, vol. 2, no. 1, pp. 4, 2017.

\bibitem{Abbe2018}
Emmanuel Abbe,
\newblock ``Community detection and stochastic block models: Recent
  developments,''
\newblock {\em Journal of Machine Learning Research}, vol. 18, no. 177, pp.
  1--86, 2018.

\bibitem{Wai2018}
Hoi-To Wai, Santiago Segarra, Asuman~E Ozdaglar, Anna Scaglione, and Ali
  Jadbabaie,
\newblock ``Community detection from low-rank excitations of a graph filter,''
\newblock in {\em 2018 IEEE International Conference on Acoustics, Speech and
  Signal Processing (ICASSP)}. IEEE, 2018, pp. 4044--4048.

\bibitem{Wai2018a}
Hoi-To Wai, Santiago Segarra, Asuman~E Ozdaglar, Anna Scaglione, and Ali
  Jadbabaie,
\newblock ``Blind community detection from low-rank excitations of a graph
  filter,''
\newblock {\em arXiv preprint arXiv:1809.01485}, 2018.

\bibitem{Schaub2018}
Michael~T. Schaub, Santiago Segarra, and John Tsitsiklis,
\newblock ``Blind identification of stochastic block models from dynamical
  observations,''
\newblock {\em in preparation}, 2018.

\bibitem{Merris1994}
Russell Merris,
\newblock ``Laplacian matrices of graphs: a survey,''
\newblock {\em Linear algebra and its applications}, vol. 197, pp. 143--176,
  1994.

\bibitem{Olfati-Saber2007}
Reza Olfati-Saber, J~Alex Fax, and Richard~M Murray,
\newblock ``Consensus and cooperation in networked multi-agent systems,''
\newblock {\em Proceedings of the IEEE}, vol. 95, no. 1, pp. 215--233, 2007.

\bibitem{Masuda2017}
Naoki Masuda, Mason~A Porter, and Renaud Lambiotte,
\newblock ``Random walks and diffusion on networks,''
\newblock {\em Physics Reports}, 2017.

\bibitem{segarra_optimal_2017}
Santiago Segarra, Antonio~G Marques, and Alejandro Ribeiro,
\newblock ``Optimal graph-filter design and applications to distributed linear
  network operators,''
\newblock {\em IEEE Transactions on Signal Processing}, vol. 65, no. 15, pp.
  4117--4131, Aug 2017.

\bibitem{damoiseaux2006consistent}
Jessica~S Damoiseaux, Serge~A Rombouts, Frederik Barkhof, Philip Scheltens,
  Cornelis~J Stam, Stephen~M Smith, and Christian~F Beckmann,
\newblock ``Consistent resting-state networks across healthy subjects,''
\newblock {\em Proceedings of the national academy of sciences}, vol. 103, no.
  37, pp. 13848--13853, 2006.

\bibitem{vershynin12}
Roman Vershynin,
\newblock ``How close is the sample covariance matrix to the actual covariance
  matrix?,''
\newblock {\em Journal of Theoretical Probability}, vol. 25, no. 3, pp.
  655--686, 2012.

\bibitem{Hoffmann2018}
Till Hoffmann, Leto Peel, Renaud Lambiotte, and Nick~S Jones,
\newblock ``Community detection in networks with unobserved edges,''
\newblock {\em arXiv preprint arXiv:1808.06079}, 2018.

\end{thebibliography}

\end{document}